\documentclass[
    ,final            
  ]
  {aipproc}

\layoutstyle{8x11single}

\begin{document}

\title{Constraining the EOS of neutron-rich nuclear matter
and properties of neutron stars with heavy-ion reactions}


\author{Bao-An Li}{
  address={Department of Physics, Texas A\&M University-Commerce, Commerce,
Texas 75429-3011, USA} }

\author{Lie-Wen Chen}{
  address={Institute of Theoretical Physics, Shanghai Jiao Tong University,
Shanghai 200240, China} }

\author{Che Ming Ko}{
  address={Cyclotron Institute and Physics Department, Texas A\&M University,
College Station, Texas 77843-3366, USA}}

\author{Plamen G. Krastev}{
  address={Department of Physics, Texas A\&M University-Commerce, Commerce,
Texas 75429-3011, USA},altaddress={Department of Physics, San Diego
State University, 5500 Campanile Drive, San Diego CA 92182-1233,
USA} }

\author{De-Hua Wen}{
  address={Department of Physics, Texas A\&M University-Commerce, Commerce,
Texas 75429-3011, USA},altaddress={Department of Physics, South
China University of Technology, Guangzhou 510641, China} }

\author{Aaron Worley}{
  address={Department of Physics, Texas A\&M University-Commerce, Commerce,
Texas 75429-3011, USA} }

\author{Zhigang Xiao}{
address={Department of Physics, Tsinghua University, Beijing 100084,
China}}

\author{Jun Xu}{
  address={Institute of Theoretical Physics, Shanghai Jiao Tong University, Shanghai 200240,
China}, altaddress={Cyclotron Institute and Physics Department,
Texas A\&M University, College Station, Texas 77843-3366, USA}}

\author{Gao-Chan Yong}{
  address={Department of Physics, Texas A\&M University-Commerce, Commerce,
Texas 75429-3011, USA},altaddress={Institute of Modern Physics,
Chinese Academy of Sciences, Lanzhou 730000, China }}

\author{Ming Zhang}{
address={Department of Physics, Tsinghua University, Beijing 100084,
China}}

\begin{abstract}
Heavy-ion reactions especially those induced by radioactive beams
provide useful information about the density dependence of the
nuclear symmetry energy, thus the Equation of State of neutron-rich
nuclear matter, relevant for many astrophysical studies. The latest
developments in constraining the symmetry energy at both sub- and
supra-saturation densities from analyses of the isopsin diffusion
and the $\pi^-/\pi^+$ ratio in heavy-ion collisions using the IBUU04
transport model are discussed. Astrophysical ramifications of the
partially constrained symmetry energy on properties of neutron star
crusts, gravitational waves emitted by deformed pulsars and the
w-mode oscillations of neutron stars are presented briefly.
\end{abstract}

\maketitle

\section{Constraining the density dependence of nuclear symmetry
energy with heavy-ion collisions}

To determine the density dependence of the nuclear symmetry energy
$E_{sym}(\rho)$, thus the equation of state (EOS) of neutron-rich
nuclear matter, has been a longstanding goal of both nuclear physics
and astrophysics. The $E_{sym}(\rho)$ is critical for understanding
not only the structure of rare isotopes and heavy-ion reactions
~\cite{LiBA98,LiBA01b,Dan02a,Bar05,LCK08}, but also many interesting
issues in astrophysics~\cite{Lat00,Lat04,Lat07,Ste05}. In this
contribution we first summarize some recent progress in constraining
the $E_{sym}(\rho)$ from analyzing the isospin diffusion and
$\pi^-/\pi^+$ ratio in heavy-ion collisions within an isospin and
momentum dependent transport model IBUU04~\cite{IBUU04}. We will
then discuss some astrophysical ramifications of the partially
constrained $E_{sym}(\rho)$.
\begin{figure}[htb]
\centering{
\includegraphics[scale=0.635]{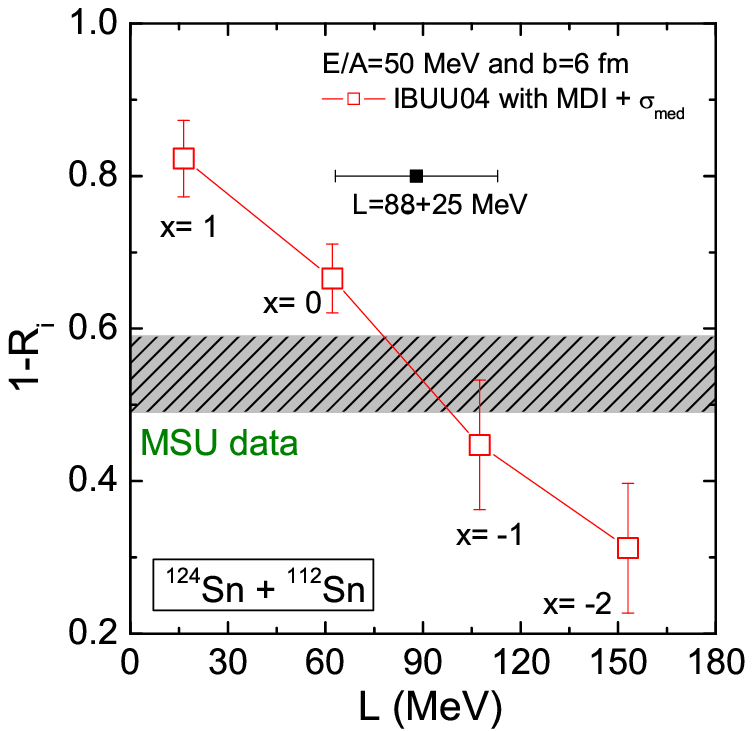}
\includegraphics[scale=0.6]{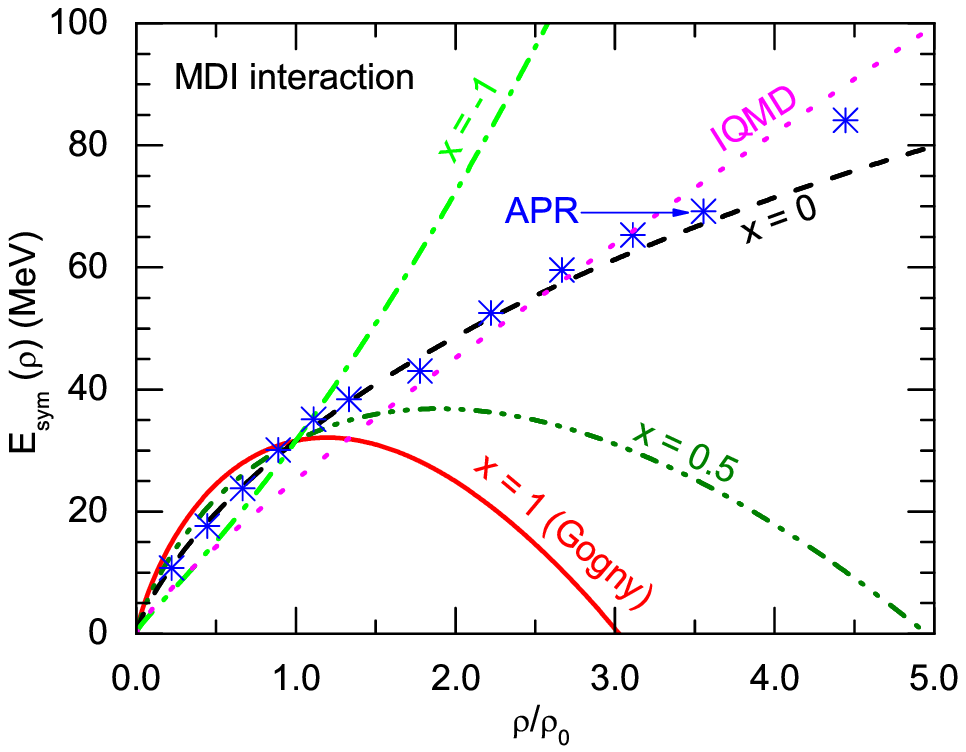}
\includegraphics[scale=0.31]{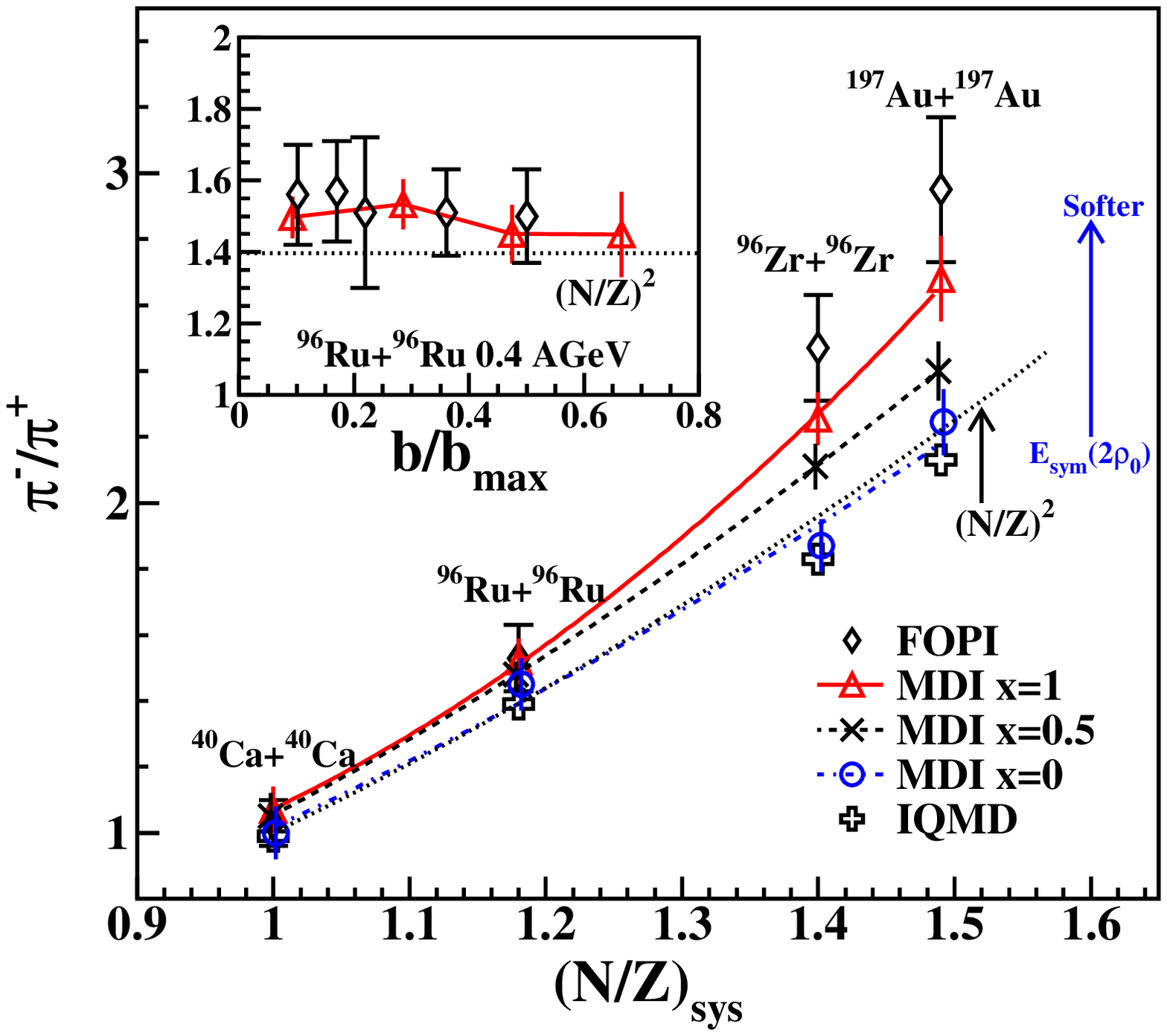}}
\caption{Left: Degree of the isospin diffusion $1-R_{i}$ as a
function of the slope parameter $L$ of the symmetry energy. Middle:
The density dependence of the symmetry energy. Right: The
$\pi^-/\pi^+$ ratio as a function of the neutron/proton ratio of the
reaction system at 0.4 AGeV with the reduced impact parameter of
$b/b_{max}\leq 0.15$. The inset is the impact parameter dependence
of the $\pi^-/\pi^+$ ratio for the $^{96}$Ru+$^{96}$Ru reaction at
0.4 AGeV. Taken from Refs.~\cite{Che05b,Xiao09}.} \label{esym00}
\end{figure}
Some interesting information about the $E_{sym}(\rho)$ has been
obtained over the last few years from heavy-ion
collisions~\cite{LCK08}. As an example, shown in the middle of
Fig.~\ref{esym00} are the $E_{sym}(\rho)$ used in the IBUU04
transport model. The three MDI $E_{sym}(\rho)$~\cite{Das03} are
obtained using a modified Gogny force by adjusting the parameter $x$
introduced in the interaction. For comparisons, the IQMD~\cite{IQMD}
and the APR\cite{Akm98} predictions are also shown. Shown in the
left window is the degree of isospin diffusion
$1-R_{i}$~\cite{Tsa04,Chen05,Liba05} versus the slope parameter
$L\equiv 3\rho _{0}\frac{\partial E_{\rm{sym}}(\rho )}{\partial \rho
}|_{\rho=\rho _{0}}$ of the MDI symmetry energy. Within the IBUU04
model analysis, the MSU data of Tsang et al.~\cite{Tsa04} favors a
$E_{sym}(\rho)$ between that with $x=0$ and $x=-1$ around the APR
prediction. More quantitatively, one can infer that the slope
parameter is about $L=88\pm 25$ MeV~\cite{Che05b}. We notice that a
very recent re-analysis of the MSU isospin diffusion data using the
ImQMD model found that the extracted range of the $E_{sym}(\rho)$
overlaps with that from the IBUU04 analysis around the
$E_{sym}(\rho)$ with $x=0$ (or APR)~\cite{Tsa08}. In the MSU isospin
diffusion reactions the maximum density reached is about
$1.2\rho_0$~\cite{Chen05}. Moreover, the isospin diffusion was found
to probe the symmetry energy during the expansion phase of the
reaction~\cite{Tsa04,Chen05}. Thus the isospin diffusion data
provides us information about the $E_{sym}(\rho)$ in the
sub-saturation density region.

Very interestingly, a recent IBUU04 transport model analysis of the
FOPI/GSI $\pi^-/\pi^+$ ratio data~\cite{Rei07} in relativistic
heavy-ion collisions at SIS/GSI with beam energies above 400 MeV/A
provides circumstantial evidence suggesting a rather soft nuclear
symmetry energy at $\rho\geq 2\rho_0$ compared to the APR
prediction~\cite{Xiao09}. Shown in the right window are the
calculated $\pi^-/\pi^+$ ratios in comparison with the FOPI data at
0.4 AGeV with the reduced impact parameter $b_0\equiv b/b_{max}\leq
0.15$ as a function of the neutron/proton ratio of the reaction
system. The inset shows the $\pi^-/\pi^+$ ratio as a function of
$b_0$ for the $^{96}$Ru+$^{96}$Ru reaction at 0.4 AGeV. It is seen
that both the data and the calculations exhibit very weak $b_0$
dependence for the $\pi^-/\pi^+$ ratio, even for mid-central
reactions where we found that the multiplicities of both $\pi^-$ and
$\pi^+$ vary appreciably with the $b_0$. For the symmetric
$^{40}$Ca+$^{40}$Ca and the slightly asymmetric $^{96}$Ru+$^{96}$Ru
reactions, calculations using both $x=1$ and $x=0$ can well
reproduce the FOPI data. Most interestingly, for the more
neutron-rich reactions of $^{96}$Zr+$^{96}$Zr and
$^{197}$Au+$^{197}$Au calculations with $x=0, 0.5$ and $1$ are
clearly separated from each other. The FOPI data favors clearly the
calculation with $x=1$. The corresponding symmetry energy is much
softer than the APR prediction. A detailed study on the excitation
function of the $\pi^-/\pi^+$ ratio also indicates that the FOPI
data can be approximately reproduced only with the $x=1$ symmetry
energy~\cite{Xiao09}. It is also interesting to mention that IQMD
calculations~\cite{Rei07} give similar results as the IBUU04 with
$x=0$. This is not surprising since the symmetry energy functionals
used in the IQMD and the IBUU04 with x=0 are very similar for
$\rho_0 <\rho \leq 3\rho_0$ as shown in the middle of
Fig.~\ref{esym00}. The most importance influence of a rather soft
symmetry energy at supra-saturation densities, such as that with
$x=1$, on pion production is through the rather high neutron/proton
ratio reached in the participant region due to the dynamical isospin
fractionation. As shown in the middle window, with $x=1$ the
$E_{sym}(\rho)$ at $\rho\geq 2\rho_0$ reached in the reaction is
very small. Thus a rather high $N/Z$ in the participant region is
energetically favored due to the dynamical isospin
fractionation~\cite{LiBA00,LiBA02} and thus the larger $\pi^-/\pi^+$
ratio is observed. In the reactions considered at 400 MeV/A, the
maximum central density reached is about $2.5\rho_0$. By varying
separately the $E_{sym}(\rho)$ at sub- and supra-saturation
densities used in the IBUU04 simulations for these reactions it was
found that the $\pi^-/\pi^+$ ratio is much more sensitive to the
variation of the $E_{sym}(\rho)$ at supra-saturation rather than
sub-saturation densities.

Putting together the information from analyzing both the isospin
diffusion and the $\pi^-/\pi^+$ ratio data, we expect that the
$E_{sym}(\rho)$ reaches a maximum somewhere between $\rho_0$ and
$2\rho_0$ before it starts decreasing at higher densities. This
indicates the importance of mapping out the $E_{sym}(\rho)$ at
densities from about $\rho_0$ to $2.5\rho_0$. Such experiments are
being planned at several facilities. If the $E_{sym}(\rho)$ is
confirmed by more experimental and theoretical studies to decrease
with increasing density above certain density, it not only posts a
serious challenge to some nuclear many-body theories but also has
important implications on several critical issues in nuclear
astrophysics, such as, the cooling of proto-neutron stars, the
possible formation of polarons due to the isospin separation
instability~\cite{Kut94,Szm06}, the possible formation of quark
droplets~\cite{Kut00} and hyperons~\cite{Ban00} in the core of
neutron stars.

Since the $E_{sym}(\rho)$ is only partially constrained in some
density regions by the available experimental data within the
IBUU04 transport model analyses, astrophysical applications of
these constraints involve some interpolations and/or
extrapolations under some assumptions. We notice here, however,
that the extrapolation of any low-density symmetry energy to
supra-saturation densities can be very dangerous as illustrated in
Fig. 1. Using the same IBUU04 transport model, at sub-saturation
densities the isospin diffusion analyses favors clearly the APR
prediction, but above about $2\rho_0$ the $\pi^-/\pi^+$ ratio data
favors instead the Gogny prediction that is much softer than the
APR prediction. Fortunately, while many astrophysical questions
depend on the symmetry energy over the whole density range, some
issues, such as the core-crust transition density in neutron stars
depends only on the $E_{sym}(\rho)$ at sub-saturation densities.
In the following section, we present several examples illustrating
the astrophysical importance of the $E_{sym}(\rho)$ using the MDI
EOS with $x=0$ and $x=-1$ extrapolated to supra-saturation
densities. Effects of the softening of the $E_{sym}(\rho)$, such
as that with $x=1$, is currently under investigation.

\section{Nuclear constraints on properties of neutron star crusts}

Neutron stars are expected to have a solid inner crust surrounding a
liquid core. Knowledge on properties of the crust plays an important
role in understanding many astrophysical
observations~\cite{Lat00,Lat04,Lat07,Ste05,BPS71,BBP71,Pet95a,Pet95b,Lin99,Hor04,Bur06,Owe05}.
The inner crust spans the region from the neutron drip-out point to
the inner edge separating the solid crust from the homogeneous
liquid core. While the neutron drip-out density $\rho _{out}$ is
relatively well determined to be
about $4\times 10^{11}$ g/cm$^{3}$ \cite{Rus06}, the transition density $%
\rho _{t}$ at the inner edge is still largely uncertain mainly
because of our very limited knowledge on the EOS, especially the
density dependence of the symmetry energy, of neutron-rich nucleonic
matter~\cite{Lat00,Lat07}. Recently, using the equation of state for
neutron-rich nuclear matter constrained by the isospin diffusion
data from heavy-ion reactions in the same sub-saturation density
range as the neutron star crust, Xu et al. put a tight constraint on
the location of the inner edge of neutron star crusts~\cite{Xu09}.

\subsection{The density and pressure at the core-crust transition}

The inner edge corresponds to the phase transition from the
homogeneous matter at high densities to the inhomogeneous matter at
low densities. In principle, the inner edge can be located by
comparing in detail relevant properties of the nonuniform solid
crust and the uniform liquid core mainly consisting of neutrons,
protons and electrons ($npe$ matter). However, this is practically
very difficult since the inner crust may contain nuclei having very
complicated geometries, usually known as the `nuclear
pasta'~\cite{Lat04,Hor04,Rav83,Oya93,Ste08}. Furthermore, the
core-crust transition is thought to be a very weak first-order phase
transition and model calculations lead to very small density
discontinuities at the transition~\cite{Pet95b,Dou00,Dou01,Hor03}.
In practice, therefore, a good approximation is to search for the
density at which the uniform liquid first becomes unstable against
small amplitude density fluctuations with clusterization. This
approximation has been shown to produce very small error for the
actual core-crust transition density and it would yield the exact
transition density for a second-order phase
transition~\cite{Pet95b,Dou00,Dou01,Hor03}. Several such methods
including the dynamical
method~\cite{BPS71,BBP71,Pet95a,Pet95b,Dou00,Oya07,Duc07}, the
thermodynamical method~\cite{Lat07,Kub07,Wor08} and the random phase
approximation (RPA)~\cite{Hor03,Hor01} have been applied extensively
in the literature.

\begin{figure}[htb]
\begin{minipage}{16.5pc}
\begin{center}
\includegraphics[scale=0.6]{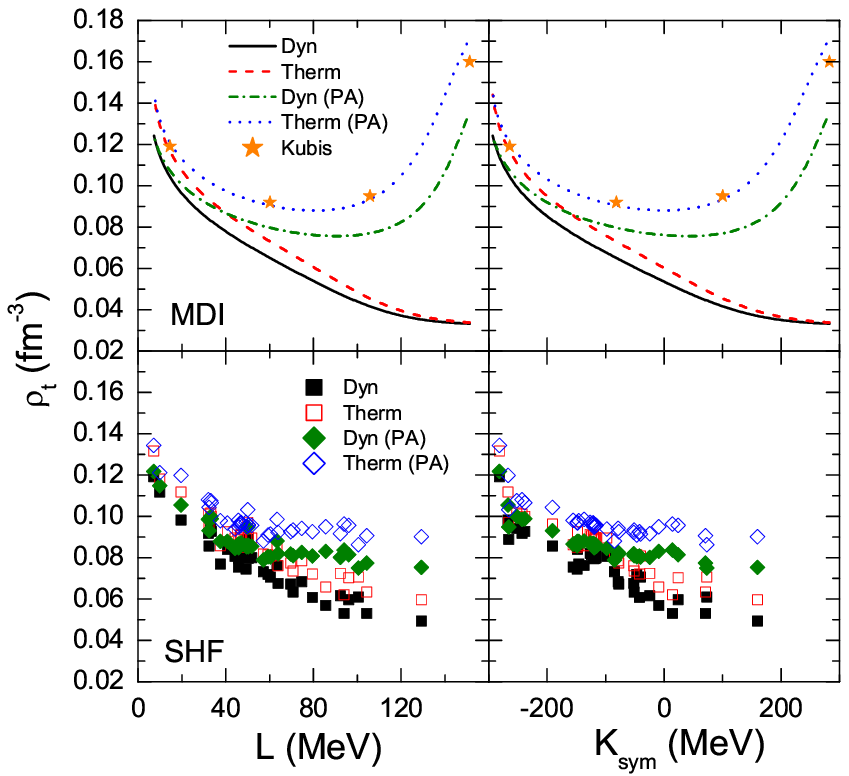}
\caption{} \label{rhotLK}
\end{center}
\end{minipage}\hspace{0pc}%
\begin{minipage}{16.5pc}
\begin{center}
\includegraphics[scale=0.6]{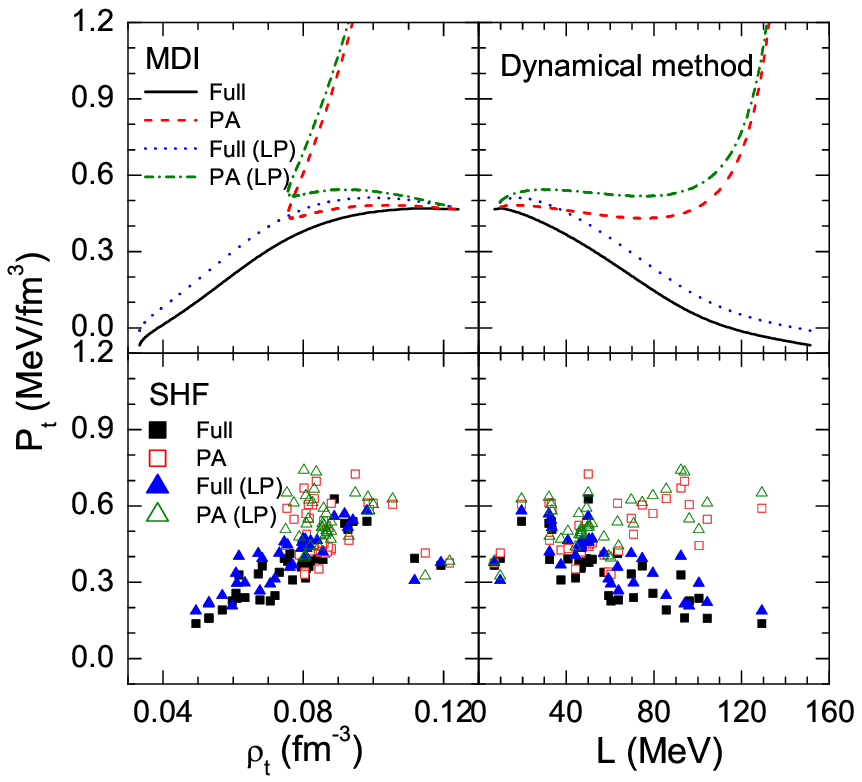}
\caption{Left: The transition density $\rho_t$ as a function of $L$
(left windows) and $K_{sym}$ (right windows) by using both the
dynamical and thermodynamical methods with the full EOS and its
parabolic approximation. Right: The transition pressure $P_t$ as a
function of $\rho_t$ and $L$ within the dynamical method with the
full EOS and its parabolic approximation. The MDI (upper windows)
and Skyrme interactions (lower windows) are used. Taken from
Ref.~\cite{Xu09}.} \label{rhotPt}
\end{center}
\end{minipage}\hspace{0pc}%
\end{figure}

Shown in the left panel of Fig.~\ref{rhotPt} is the $\rho_t$ as a
function of $L$ and $K_{sym}\equiv 9\rho^2_0\frac{\partial^2
E_{sym}(\rho)}{\partial \rho^2}|_{\rho=\rho_0}$ using both the
dynamical and thermodynamical methods with the full EOS and its
parabolic approximation (PA) from the MDI interaction with the
varying $x$ parameter and 47 Skyrme forces~\cite{Xu09}. With the
full MDI EOS, it is clearly seen that the $\rho _{t}$ decreases
almost linearly with increasing $L$ for both methods. This feature
is consistent with the RPA results~\cite{Hor01} and that found
recently by Oyamatsu et al.~\citep{Oya07}. The similar relation is
also observed between the $\rho_t$ and $K_{sym}$ due to the fact
that $K_{sym}$ always correlate with $L$ for a fixed energy density
functional~\cite{Xu09}. It is interesting to see that both the
dynamical and thermodynamical methods give very similar results. On
the other hand, surprisingly, the PA drastically changes the
results, especially for stiffer symmetry energies (larger $L$
values). Also included in the left panel of Fig.~\ref{rhotPt} are
the predictions by Kubis using the PA of the MDI EOS in the
thermodynamical approach~\cite{Kub07}. The large error introduced by
the PA is understandable since the $\beta $-stable $npe$ matter is
usually highly neutron-rich and the contribution from the higher
order terms in $\delta $ is appreciable. This is especially the case
for the stiffer symmetry energy which generally leads to a more
neutron-rich $npe $ matter at subsaturation densities. In addition,
simply because of the energy curvatures involved in the stability
conditions, the contributions from higher order terms in the EOS are
multiplied by a larger factor than the quadratic term. These
features agree with the early finding~\cite{Arp72} that the $\rho
_{t}$ is very sensitive to the fine details of the nuclear EOS.
Applying the experimentally constrained $L$ values of $86\pm25$ MeV
to the $\rho_t-L$ correlation obtained using the full EOS within the
dynamical method shown in the left panel of Fig.~\ref{rhotPt}, one
can conclude that the transition density is between $0.040$
fm$^{-3}$ and $0.065$ fm$^{-3}$. This constrained range is
significantly below the fiducial value of $\rho_t\approx 0.08$
fm$^{-3}$ often used in the literature and the estimate of
$0.5<\rho_t/\rho_0<0.7$ made in ref.~\citep{Lat07} within the
thermodynamical approach using the parabolic approximation of the
EOS.

The pressure at the inner edge is an important quantity related
directly with the crustal fraction of the moment of inertia which
can be measurable indirectly from observations of pulsar
glitches~\cite{Lat07}. It is very instructive to quote the
analytical estimation obtained by Lattimer and Prakash~\cite{Lat07}
for the transition pressure
\begin{eqnarray}\label{lp}
P_t &=&
\frac{K_0}{9}\frac{\rho_t^2}{\rho_0}\left(\frac{\rho_t}{\rho_0}-1\right)
+\rho_t\delta_t
\left[\frac{1-\delta_t}{2}E_{sym}(\rho_t)+\left(\rho\frac{d
E_{sym}(\rho)}{d \rho}\right)_{\rho_t}\delta_t\right],
\end{eqnarray}
where $K_0$ is the incompressibility of symmetric nuclear matter at
$\rho_0$ and $\delta_t$ is the isospin asymmetry at $\rho_t$. One
can see that, besides the implicit dependence on the symmetry energy
through the $\rho_t$ and $\delta_t$, the $P_t$ also depends
explicitly on the value and slope of the $E_{sym}(\rho)$ at
$\rho_t$. Thus the $P_t$ depends very sensitively on the
$E_{sym}(\rho)$. Shown in the right panel of Fig.~\ref{rhotPt} is
the $P_t$ as a function of $\rho_t$ (left windows) and $L$ (right
windows) by using the dynamical method with and without the
parabolic approximation. The results from Eq.~(\ref{lp}) using the
$\rho_t$ and $E_{sym}$ corresponding to the full EOS and its PA are
also shown for comparisons. It is interesting to see that the
Eq.~(\ref{lp}) predicts qualitatively the same but quantitatively
slightly higher values compared to the original expressions for the
pressure with or without the PA for both the thermodynamical and
dynamical methods even though this formula was derived from the
thermodynamical method using the PA. This observation implies that
the direct effect of using the full EOS or its PA on the pressure is
small although the PA may affect strongly the transition pressure
$P_t$ by changing the transition density $\rho_t$. The $P_t$
essentially increases with the increasing $\rho_t$ in calculations
using the full EOS, but a complex relation between the $P_t$ and
$\rho_t$ is obtained using the PA. The large difference in $P_t$ is
due to the strong PA effect on the $\rho_t$. Moreover, the latter
does not vary monotonically with $L$ for the PA as shown in the
right panel of Fig.~\ref{rhotPt}.

\begin{figure}[t!]
\centering
\includegraphics[scale=0.6]{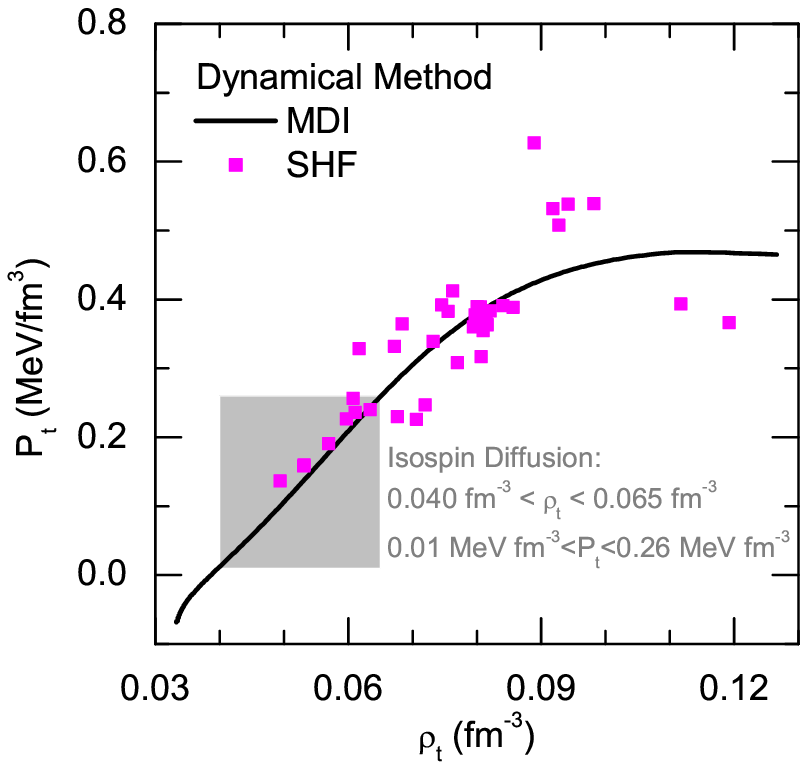}
\caption{$P_t$ as a function of $\rho_t$ by using the dynamical
method without parabolic approximation for both MDI interaction and
SHF calculations. The shaded band represent the constraint from the
isospin diffusion data. Taken from Ref.~\cite{Xu09}.}
\label{PtRhotDyn}
\end{figure}
It is also interesting to examine the range of $P_t$ corresponding
to the $\rho_t$ and $L$ constrained by the heavy-ion reaction data.
In Fig.~\ref{PtRhotDyn}, we show the $P_t$ as a function of $\rho_t$
by using the dynamical method and the full EOS for both the MDI
(solid line) and the Skyrme (filled squares) calculations. It is
interesting to see that the MDI and Skyrme interactions give
generally quite consistent results. Corresponding to the $\rho_t$
constrained in between 0.040 fm$^{-3}$ and 0.065 fm$^{-3}$, the
$P_{t}$ is limited between $0.01$ MeV/fm$^{3}$ and $0.26$
MeV/fm$^{3}$ with the MDI interaction as indicated by the shaded
area, which is significantly less than the fiducial value of
$P_t\approx 0.65$ MeV/fm$^{3}$ often used in the literature
~\cite{Lat07}. As pointed out in a recent work by Avancini et
al~\cite{Ava08b}, the value of $P_t\approx 0.65$ MeV/fm$^{3}$ may be
too large for most mean-field calculations without the PA. We note
that among the $47$ Skyrme interactions used here, the following $8$
interactions, i.e., the SkMP, SKO, R$_\sigma$, G$_\sigma$, SV, SkI2,
SkI3, and SkI5, are consistent with the constraints from heavy-ion
reactions. These results indicate that one may introduce a huge
error by assuming {\it a priori} that the EOS is parabolic for a
given interaction in calculating the $\rho _{t}$ and  $P_{t}$.

\subsection{The core size and crust thickness}

For a given neutron star of total mass $M$ and radius $R$, what are
the respective sizes of its core and crust? How do they depend on
the stiffness of the symmetry energy? How do they depend on the
neutron star mass $M$?  How does the thickness of neutron star
crusts depend on $L$ while it is well known that the size of neutron
skin in heavy nuclei increases with the increasing
$L$~\cite{Che05b}? These questions have been investigated recently
in Ref.~\cite{Xu09}.

\begin{figure}[t!]
\centering
\includegraphics[scale=0.65]{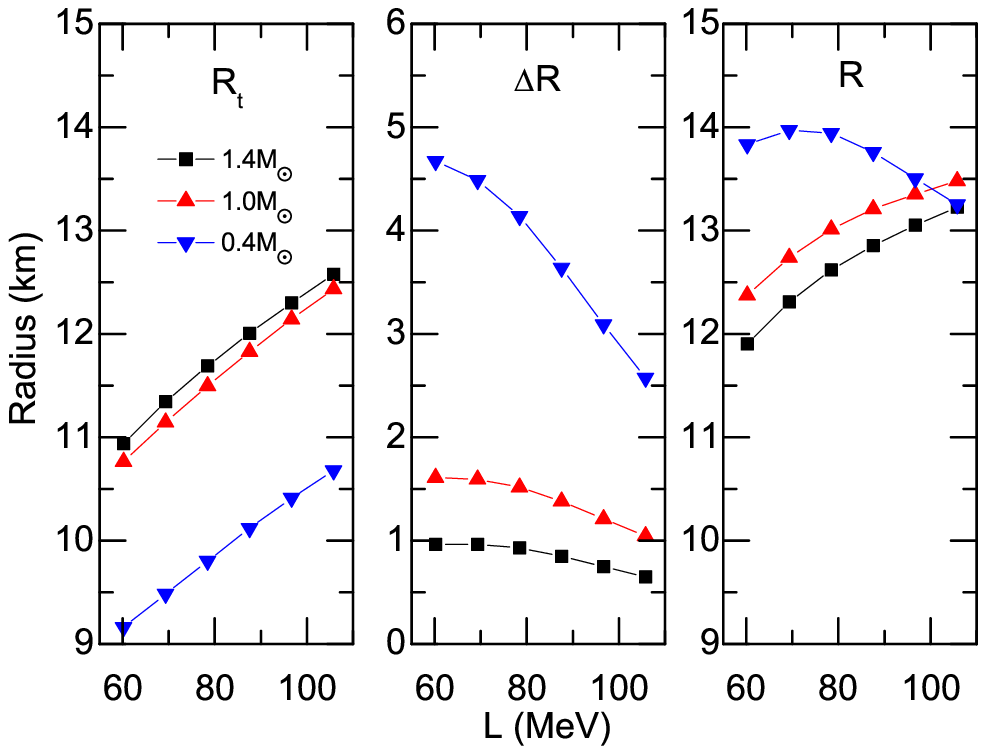}
\includegraphics[scale=0.65]{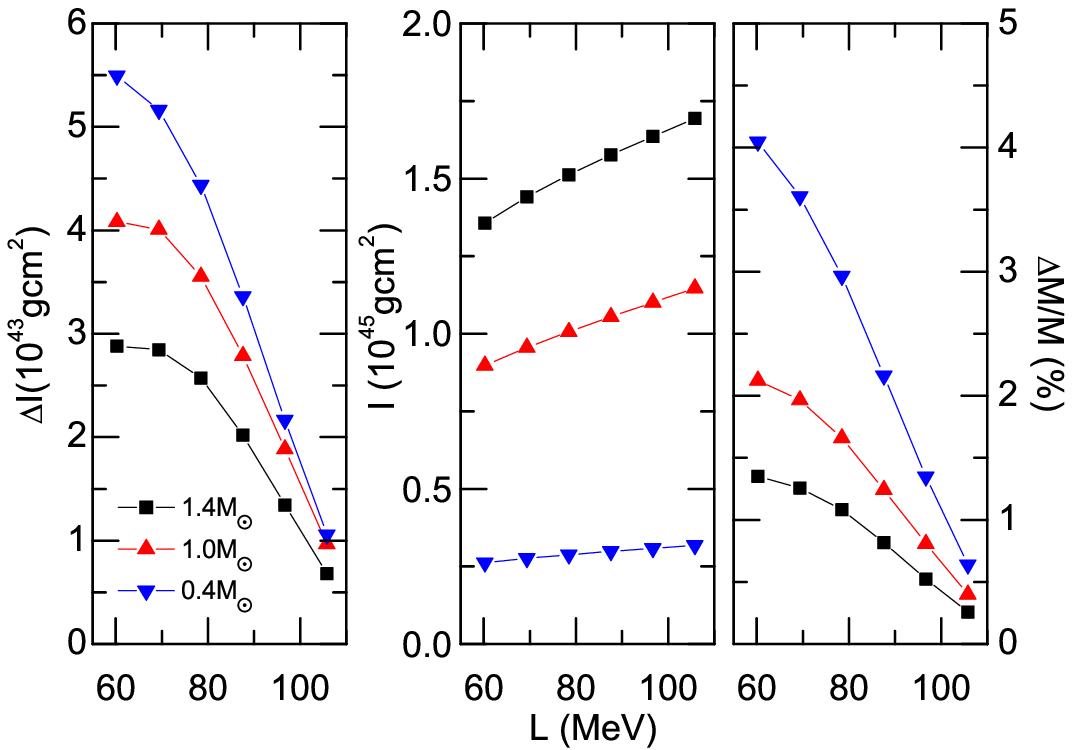}
\caption{{\protect\small Left: The whole radius $R$, the crust
thickness $\Delta R$, the core radius $R_t$ as functions of $L$ at
fixed total mass of $0.4M_{\odot}$, $1.0M_{\odot}$ and
$1.4M_{\odot}$, respectively. Right: The crustal fraction of neutron
mass $\Delta M/M$, the moment of inertia $I$ of the whole star and
the crust contribution $\Delta I$ as a function of $L$, at fixed
total mass $0.4M_{\odot}$, $1.0M_{\odot}$ and $1.4M_{\odot}$,
respectively. Taken from Ref.~\cite{Xu09}.}}\label{RL}
\end{figure}

Show in left panel of Fig.~\ref{RL} are the core radius $R_t$, the
crust thickness $\Delta R$ and the total radius $R$ as functions of
$L$ for a fixed total mass of $0.4M_{\odot}$, $1.0M_{\odot}$ and
$1.4M_{\odot}$, respectively. It is seen that the $R_t$ increases
almost linearly with increasing $L$. The $R_t$ also increases with
the increasing mass at a fixed $L$. This is because the stiffer the
symmetry energy is, the larger the contribution of the isospin
asymmetric part of the pressure will be, which makes the $R_t$
larger. Moreover, the $\Delta R$ decreases with the increasing $L$
especially for light neutron stars, as the transition density
decreases with the increasing $L$. As the thickness of the crust
$\Delta R$ and the core radius $R_t$ depend oppositely on $L$, the
total radius $R=R_t+\Delta R$ of the neutron star may show a
complicated dependence on $L$. We stress here that this is the
result of a competition between the repulsive nuclear pressure
dominated by the symmetry energy contribution and the gravitational
binding. Interestingly, it is often mentioned that the crust of
neutron stars bears some analogy with the neutron-skin of heavy
nuclei. However, they show completely opposite dependences on the
$L$. Namely, the size of neutron-skin usually increases with the
increasing $L$ as a result of the competition between the nuclear
surface tension and the pressure difference of neutrons and
protons~\cite{Che05b}, while the thickness of neutron star crusts
decreases with the increasing $L$ as a result of the competition
between the nuclear pressure and the gravitational binding.

\subsection{The fractional mass and momenta of inertia of neutron star crust}

What is the crustal fraction $\Delta M/M$ of the total mass and how
does it depend on the symmetry energy? Since the mass is simply the
integration of the energy density, one expects the $\Delta M/M$ and
$\Delta R/R$ have very similar dependences on $L$~\cite{Xu09}. Shown
in the right window of Fig.~\ref{RL} (right panel) is the $\Delta
M/M$. The fractional mass of the crust decreases with the increasing
$L$ at a fixed total mass, and it decreases with the increasing
total mass at a fixed value of $L$. The moment of inertia is
determined by the distribution of the energy density. From the
middle window, it is seen that the total moment of inertia increases
with the increasing mass at a fixed value of $L$ and increases with
the increasing $L$ at a fixed total mass. The dependence on $L$ is
relatively weak especially for light neutron stars. However, the
crust contribution of the moment of inertia varies much more quickly
with $L$. It decreases with the increasing neutron star mass at a
fixed value of $L$ and decreases with the increasing $L$ at a fixed
total mass. These are all consistent with the behaviors of the
fractional mass and size of the crust.

The crustal fraction of the moment of inertia ${\Delta I}/{I}$ is
particularly interesting as it can be inferred from observations of
pulsar glitches, the occasional disruptions of the otherwise
extremely regular pulsations from magnetized, rotating neutron
stars. It can be expressed approximately as~\cite{Lat07,Lat00}
\begin{eqnarray}\label{dI}
\frac{\Delta I}{I} &=& \frac{28\pi P_t R^3}{3 M c^2}
\frac{(1-1.67\xi-0.6\xi^2)}{\xi}
\left[1+\frac{2P_t(1+5\xi-14\xi^2)}{\rho_t m c^2 \xi^2}\right]^{-1},
\end{eqnarray}
where $m$ is the mass of baryons and $\xi=G M/R c^2$. This
analytical formula has been verified by direct numerical
calculations using both the full EOS and its PA~\cite{Xu09}.
Furthermore, it is indicated that there exists big differences for
$\Delta I/I$ by comparing calculations using the full EOS and its
PA~\cite{Xu09}, again due to the corresponding differences in the
transition density. As it was stressed in ref.~\cite{Lat00}, the
$\Delta I/I$ depends sensitively on the symmetry energy at
sub-saturation densities through the $P_t$ and $\rho_t$, but there
is no explicit dependence upon the EOS at higher-densities.
Experimentally, the crustal fraction of the moment of inertia has
been constrained as ${\Delta I}/{I}>0.014$ from studying the
glitches of the Vela pulsar~\cite{Lin99}. Combining the
observational constraint of $\Delta I/I>0.014$ with the upper bounds
of $\rho _{t}=0.065$ fm$^{-3}$ and $P_{t}=0.26$ MeV/fm$^{3}$
inferred from heavy-ion reactions, we can obtain a minimum radius of
$R\geq 4.7+4.0M/M_{\odot }$ km for the Vela pulsar. According to
this constraint, the radius of the Vela pulsar is predicted to
exceed $10.5$ km should it have a mass of $1.4M_{\odot }$. We notice
that a constraint of $R\geq 3.6+3.9M/M_{\odot }$ km for this pulsar
has previously been derived in Ref.~\cite{Lin99} by using $\rho
_{t}=0.075$ fm$^{-3}$ and $P_{t}=0.65$ MeV/fm$^{3}$. However, the
constraint obtained here using data from both the terrestrial
laboratory experiments and astrophysical observations is
significantly different and actually it is more stringent.

\section{Imprints of symmetry energy on gravitation waves}
Gravitational waves are tiny disturbances in space-time and are a
fundamental, although not yet directly confirmed, prediction of
General Relativity. They can be triggered in cataclysmic events
involving (compact) stars and/or black holes. They could even have
been produced during the very early Universe, well before any stars
had been formed, merely as a consequence of the dynamics and
expansion of the Universe. Because gravity interacts extremely
weakly with matter, gravitational waves would carry a genuine
picture of their sources and thus provide undisturbed information
that no other messenger can deliver~\cite{Maggiore:2007}.
Gravitational wave astrophysics would open an entirely new
non-electromagnetic window making it possible to probe physics that
is hidden or dark to current electromagnetic
observations~\cite{Flanagan:2005yc}.

Deformed pulsars and various oscillation modes of spherical neutron
stars are among the possible sources of gravitational waves. In
particular, the deformed pulsars are major candidates for sources of
continuous gravitational waves in the frequency bandwidth of the
existing ground-based laser interferometric detectors including the
LIGO~\cite{Abbott:2004ig} and VIRGO (e.g.
Ref.~\cite{Acernese:2007zzb}). While the oscillations of neutron
stars are mostly at frequencies much higher than the currently
existing and planned gravitational wave observatories, their studies
are of fundamental theoretical interest. The strain-amplitude of
gravitational waves from deformed pulsars depends on the star's
quadrupole moment determined by the EOS of neutron-rich nuclear
matter. On the other hand, both the frequency and the decay rate of
the fundamental oscillation modes are determined also by the EOS of
neutron-rich nuclear matter. In the following we present several
examples illustrating the imprints of the symmetry energy on
gravitation waves. More details can by found in
refs.~\cite{Kra08b,Wor09,Wen09}.

\subsection{Gravitational waves from deformed pulsars}

\begin{figure}[t!]
\centering
\includegraphics[height=4.5cm]{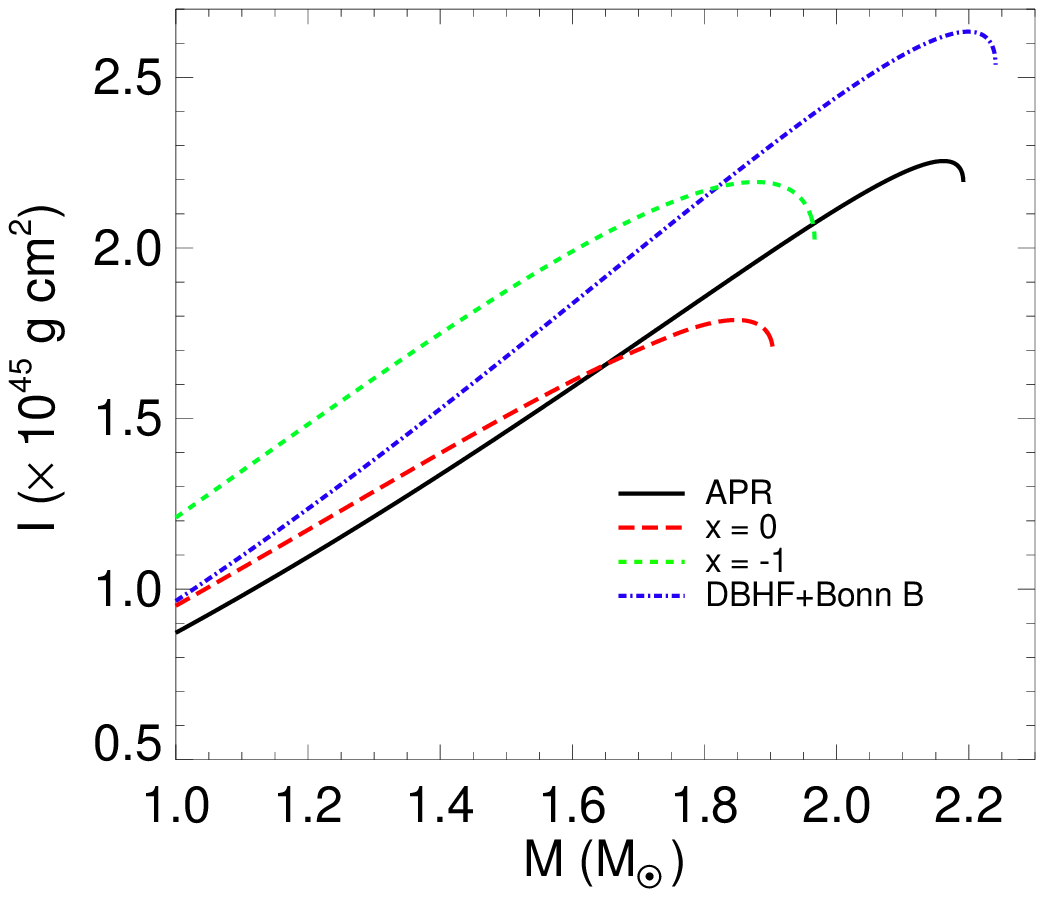}
\includegraphics[height=4.5cm]{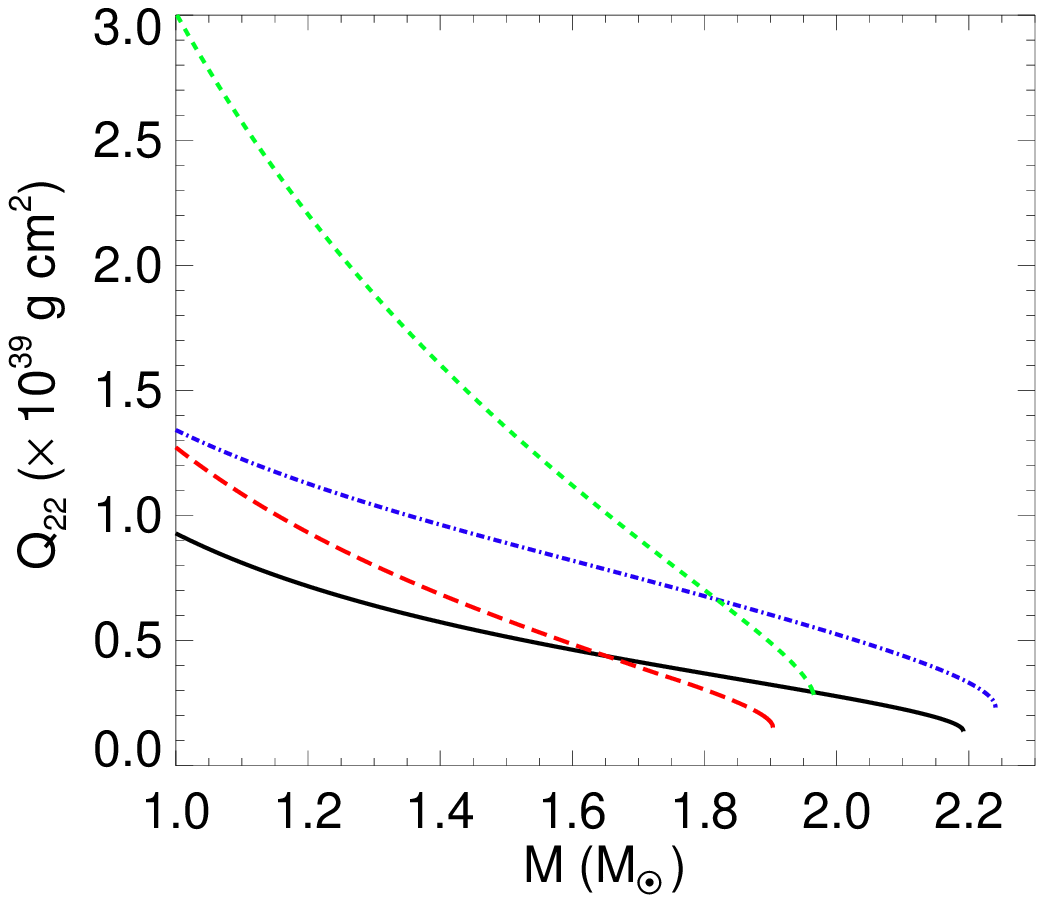}
\includegraphics[height=4.5cm]{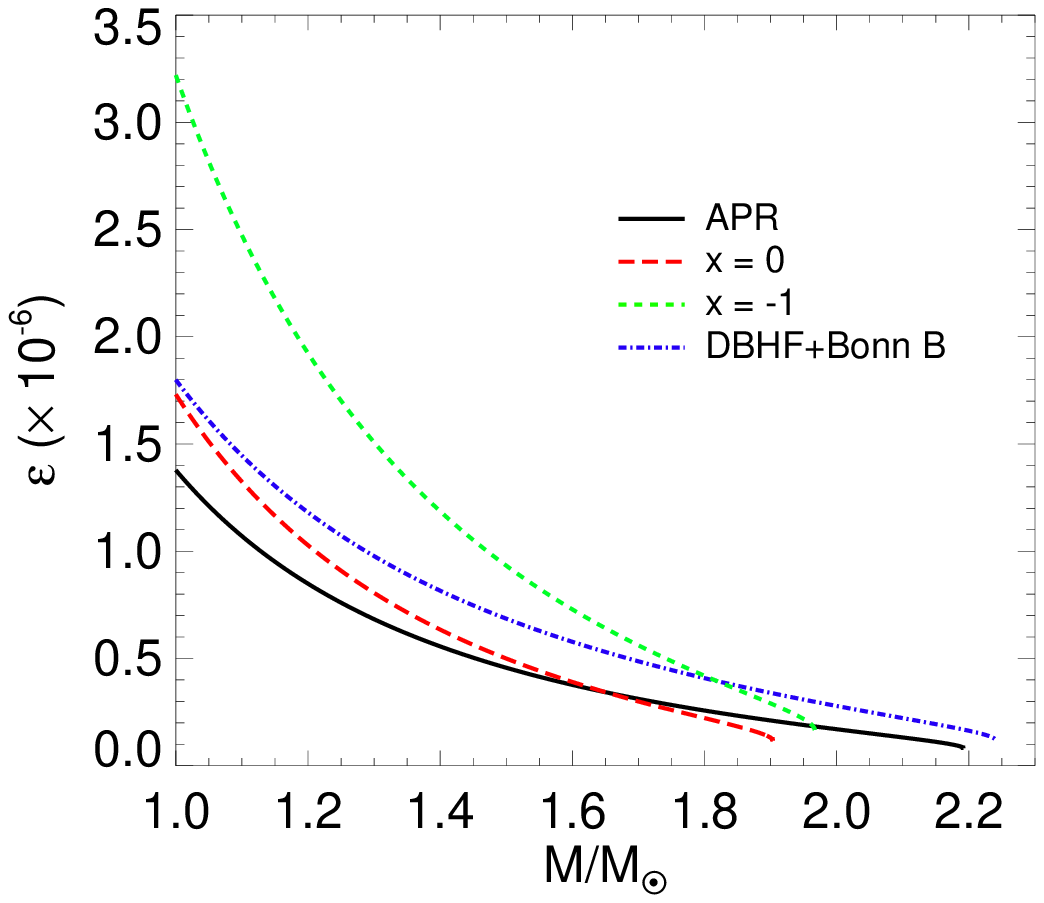}
\caption{ Neutron star moment of inertia (left), quadrupole moment
(middle) and ellipticity (right) as functions of the neutron star
mass. Taken from Ref.~\cite{Kra08b,Kra08a}} \label{GW1}
\end{figure}

The strain amplitude of gravitational waves at the Earth's
vicinity (assuming an optimal orientation of the rotation axis
with respect to the observer) from deformed pulsars can be written
as~\cite{HAJS:2007PRL}
\begin{equation}\label{Eq.1}
h_0=\frac{16\pi^2G}{c^4}\frac{\epsilon I_{zz}\nu^2}{r},
\end{equation}
where $\nu$ is the neutron star rotational frequency, $I_{zz}$ its
principal moment of inertia, $\epsilon=(I_{xx}-I_{yy})/I_{zz} $
its equatorial ellipticity, and $r$ its distance to Earth. The
ellipticity is related to the neutron star maximum quadrupole
moment (with $m=2$) via~\cite{Owen:2005PRL}
\begin{equation}\label{Eq.2}
\epsilon = \sqrt{\frac{8\pi}{15}}\frac{\Phi_{22}}{I_{zz}},
\end{equation}
where for {\it slowly} rotating (and static) neutron stars
$\Phi_{22}$ can be written as~\cite{Owen:2005PRL}
\begin{equation}\label{Eq.3}
\Phi_{22,max}=2.4\times
10^{38}g\hspace{1mm}cm^2\left(\frac{\sigma}{10^{-2}}\right)\left(\frac{R}{10km}\right)^{6.26}
\left(\frac{1.4M_{\odot}}{M}\right)^{1.2}
\end{equation}
In the above expression $\sigma$ is the breaking strain of the
neutron star crust which is rather uncertain at present time and
lies in the range $\sigma=[10^{-5}-10^{-2}]$~\cite{HAJS:2007PRL}.
From Eqs.~(\ref{Eq.1}) and (\ref{Eq.2}) it is clear that $h_0$
does not depend on the moment of inertia $I_{zz}$, and that the
total dependence upon the EOS is carried by the quadrupole moment
$\Phi_{22}$. Thus Eq.~(\ref{Eq.1}) can be rewritten as
\begin{equation}\label{Eq.4}
h_0=\chi\frac{\Phi_{22}\nu^2}{r},
\end{equation}
with $\chi=\sqrt{2045\pi^5/15}G/c^4$. For slowly rotating neutron
stars Lattimer and Schutz~\cite{Lattimer:2005} derived the
following empirical relation for the moment of inertia
\begin{equation}\label{Eq.5}
I\approx (0.237\pm
0.008)MR^2\left[1+4.2\frac{Mkm}{M_{\odot}R}+90\left(\frac{Mkm}{M_{\odot}R}\right)^4\right]
\end{equation}
Using Eq.~(\ref{Eq.5}) to calculate the neutron star moment of
inertia and Eq.~(\ref{Eq.3}) the corresponding quadrupole moment,
the ellipticity $\epsilon$ can be readily computed (via
Eq.~(\ref{Eq.2})). The results are shown in Fig.~\ref{GW1}. It is
clearly seen that the EOS, especially the symmetry energy, has a
strong effects on all of these important quantities.

Since the global properties of spinning neutron stars (in
particular the moment of inertia) remain approximately constant
for rotating configurations at frequencies up to $\sim
300Hz$~\cite{Wor08}, the above formalism can be readily employed
to estimate the gravitational wave strain amplitude, provided one
knows the exact rotational frequency and distance to Earth, and
that the frequency is relatively low (below $\sim 300Hz$).

\begin{figure}[!t]
\centering
\includegraphics[totalheight=4.5cm]{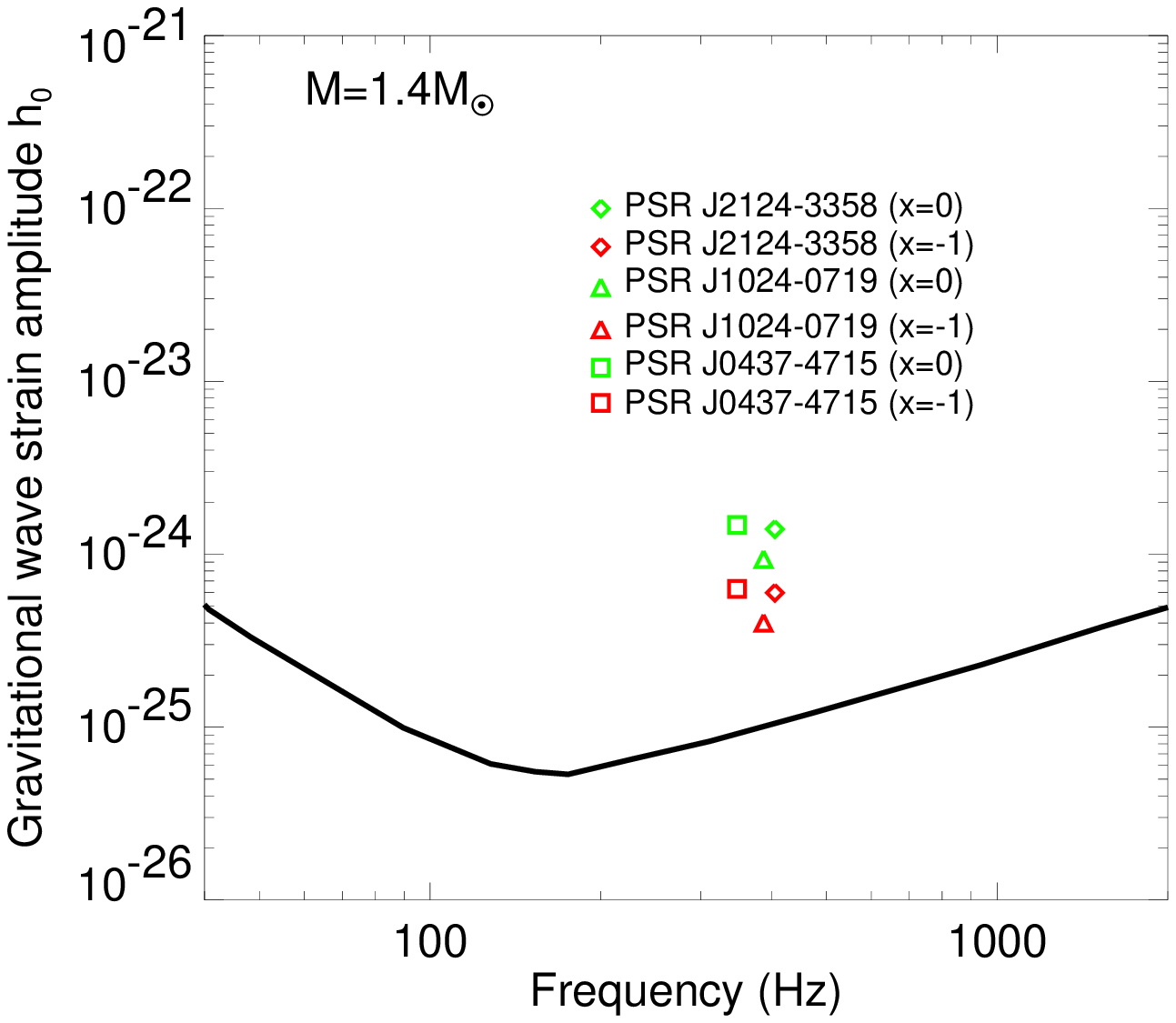}
\includegraphics[totalheight=4.5cm]{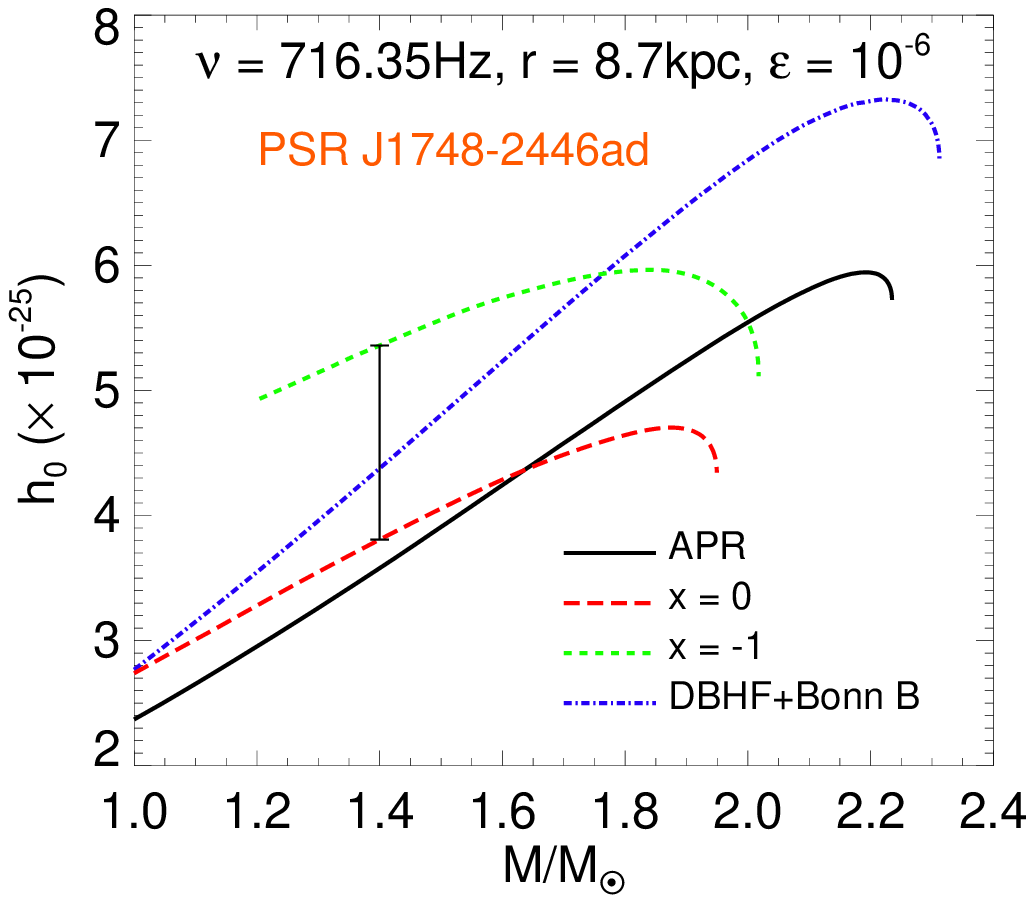}
\vspace{5mm} \caption{Gravitational-wave strain amplitude as a
function of frequency for several slow pulsars (left) and PSR
J1748-2446 (right). Taken from Ref.~\cite{Kra08b,Wor09}} \label{GW2}
\end{figure}

Shown in the left window of Fig.~\ref{GW2} is the GW strain
amplitude, $h_0$, as a function of frequency for several slowly
rotating near-earth neutron stars. At higher rotational frequencies,
relativistic rotational models have to be used~\cite{Wor09}. The
solid line represents the designed upper detection limit of LIGO.
Shown in the right window is the $h_0$ for the PSR J1748-2446
rotating at 716z assuming it has an ellipticity of
$\epsilon=10^{-6}$. In both cases clear imprints of the symmetry
energy are seen.

\subsection{{Gravitational waves from the axial w-mode oscillations of neutron stars}}
In the framework of general relativity, gravitational radiation
damps out the neutron star oscillations which leads to the frequency
of the non-radial oscillations to become "quasi-normal" (complex)
with a real part representing the actual frequency of the
oscillation and an imaginary part representing the losses due to its
damping~\cite{Cha91}. The eigen-frequencies of the quasi-normal
modes could be found by solving the equations which describe the
non-radial perturbations of a static neutron star in general
relativity. The critical input to solve the equation is the nuclear
EOS. The so called w-mode associated with the space-time
perturbation only exists in general relativity. It is very important
for astrophysical applications since it is related to the space-time
curvature and exists for all relativistic stars, including black
holes.
\begin{figure}
\includegraphics[totalheight=6cm]{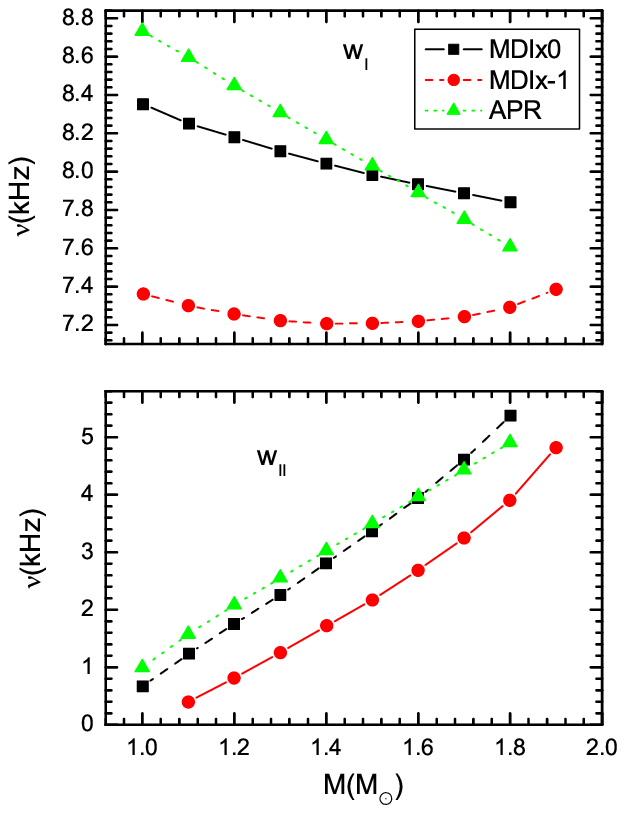}
\includegraphics[totalheight=6cm]{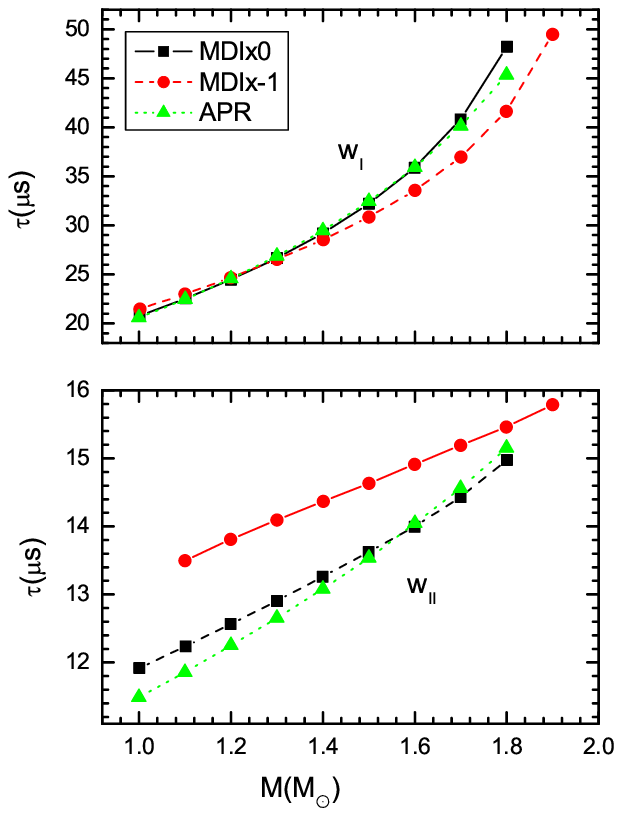}
\caption{\label{wen1}Frequency (left) and decay time (right) of
$w_{I}$-mode (upper) and $w_{II}$-modes  (lower) as functions of the
neutron star mass $M$. Taken from ref.~\cite{Wen09}}
\end{figure}
Using the MDI EOS, Wen et al. recently studied the axial
w-modes~\cite{Wen09}. Figure~\ref{wen1} displays both the frequency
and damping time of the $w_{I}$- (upper frame) and $w_{II}$-modes
(lower frame) respectively, as a function of the neutron star mass.
The results establish the relationship between the expected
frequencies of the axial w-modes, for a given EOS, and the stellar
mass. It is interesting to notice that there is a clear difference
between the frequencies calculated with the MDI ($x=0$) EOS and
those with the MDI ($x=-1$) EOS. Since the major difference between
these two cases is the density dependence of the nuclear symmetry
energy, it is obvious that the symmetry energy has a clear imprint
on the frequencies.
\begin{figure}
\includegraphics[totalheight=6cm]{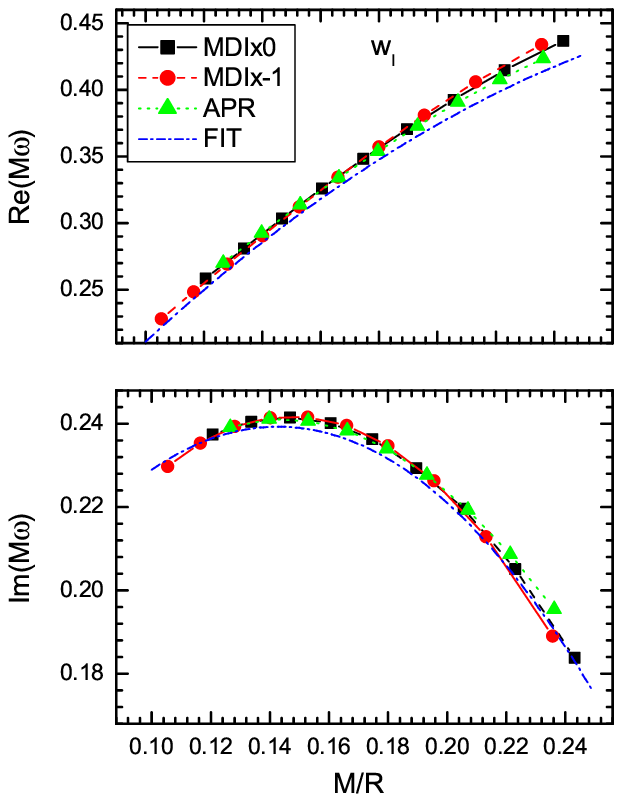}
\includegraphics[totalheight=6cm]{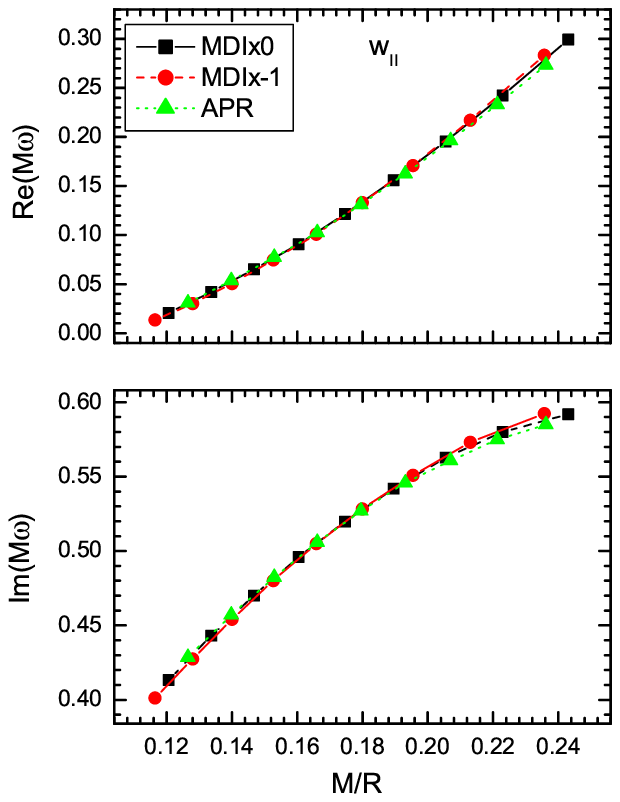}
\caption{\label{wen2} The scaled frequency (left) and decay time
(right) of $w_{I}$-mode (upper) and $w_{II}$-modes  (lower) as
functions of the neutron star compactness $M/R$. Taken from
ref.~\cite{Wen09}}
\end{figure}

Shown in Fig.~\ref{wen2} are the real (upper panel) and imaginary
(lower panel) parts of the eigen-frequency of $w_{I}$ and
$w_{II}$-modes scaled by the mass $M$ as a function of the
compactness parameter $M/R$, respectively.  These results suggest
that the scaled eigen-frequency exhibits a universal behavior as a
function of the compactness parameter independent of the EOS used.
As discussed by Andersson~\cite{Nil98}, Benhar~\cite{Ben99} and Tsui
~\cite{Tsu05} this finding could be used to constrain the frequency
and damping time of gravitational waves. This is very important for
guiding the gravitational wave search provided the mass and radius
of the prospective source (neutron star) are known. On the other
hand, when the gravitational wave astronomy becomes a reality,
namely if both the frequency and dumping time for a given neutron
star are known this could provide information on the neutron star
mass and radius. In Figs.~\ref{wen2} the fitting curves representing
the results of Tsui et al.~\cite{Tsu05} are also shown. It is seen
that the numerical results of Wen et al. are in good agreement with
those of Tsui et al.~\cite{Tsu05}.

\section{Summary}
In summary, important progress has been made in recent years in
constraining the symmetry energy with heavy-ion collisions. Their
implications in some astrophysical phenomena have been explored.
Nevertheless, the field is still at its beginning. While a number of
potentially useful probes of the $E_{sym}(\rho)$ have been proposed,
available experimental data are mostly for reactions with stable
beams~\cite{LCK08}. Coming experiments with more neutron-rich nuclei
at several advanced radioactive beam facilities are expected to
improve the situation dramatically. Thus, more exciting times are
yet to come.

\begin{theacknowledgments}
This work was supported in part by the US National Science
Foundation under Grant No. PHY-0652548, PHY-0757839, PHY-0758115 and
PHY-0457265, the Welch Foundation under Grant No. A-1358, the
Research Corporation under Award No. 7123, the Texas Coordinating
Board of Higher Education Award No. 003565-0004-2007, the National
Natural Science Foundation of China under Grant Nos. 10575071,
10675082, 10874111, and 10647116, MOE of China under project
NCET-05-0392, Shanghai Rising-Star Program under Grant No.
06QA14024, the SRF for ROCS, SEM of China, the National Basic
Research Program of China (973 Program) under Contract No.
2007CB815004, and the Young Teachers' Training Program of the
Chinese Scholarship Council under Grant No. 2007109651.
\end{theacknowledgments}


\begin{thebibliography}{9}

\bibitem{LiBA98} B.A. Li, C.M. Ko and W. Bauer, Int. Jour.
Mod. Phys. E {\bf 7}, 147 (1998).

\bibitem{LiBA01b} Isospin Physics in Heavy-Ion Collisions at Intermediate
Energies, Eds. Bao-An Li and W. Udo Schr\"{o}der (Nova Science
Publishers, Inc, New York, 2001).

\bibitem{Dan02a} P. Danielewicz, R. Lacey, W.G. Lynch, Science {\bf 298}, 1592 (2002).

\bibitem{Bar05} V. Baran, M. Colonna, V. Greco and M. DiToro, Phys. Rep. {\bf 410}, 335 (2005).

\bibitem{LCK08} B.A. Li, L.W. Chen and C.M. Ko, Phys. Rep. {\bf 464}, 113 (2008).

\bibitem{Lat00} J.M. Lattimer and M. Prakash, Phys. Rep. \textbf{333-334},
121 (2000); Astrophys. J. \textbf{550}, 426 (2001).

\bibitem{Lat04} J.M. Lattimer and M. Prakash, Science \textbf{304}, 536
(2004).

\bibitem{Lat07} J.M. Lattimer and M. Prakash, Phys. Rep. \textbf{442}, 109
(2007).

\bibitem{Ste05} A.W. Steiner, M. Prakash, J.M. Lattimer, and P.J. Ellis, Phys. Rep. \textbf{410}, 325
(2005).

\bibitem{IBUU04}B.A. Li, C.B. Das, S. Das Gupta and C. Gale, Nucl. Phys. {\bf
A735}, 563 (2004); Phys. Rev. {\bf C69}, 064602 (2004).

\bibitem{Das03}C. B. Das, S. Das Gupta, C. Gale and B.A. Li, Phys. Rev. {\bf C67}, 034611 (2003).

\bibitem{IQMD} C. Hartnack, Rajeev K. Puri, J. Aichelin, J. Konopka, S.A. Bass, H. Stoecker and W.
Greiner, Euro. Phys. J., {\bf A1}, 151 (1998).

\bibitem{Akm98} A. Akmal, V.R. Pandharipande and D.G. Ravenhall, Phys. Rev. C 58 (1998) 1804.

\bibitem{Tsa04}M.B. Tsang et al., Phys. Rev. Lett. {\bf 92}, 062701 (2004).

\bibitem{Chen05}L.W. Chen, C.M. Ko and B.A. Li Phys. Rev. Lett. {\bf 94}, 032701 (2005).

\bibitem{Liba05}B.A. Li and L.W. Chen, Phys. Rev. {\bf C72}, 064611 (2005).

\bibitem{Che05b} L.W. Chen, C.M. Ko, and B.A. Li, Phys. Rev. C \textbf{72},
064309 (2005).

\bibitem{Tsa08} M.B. Tsang et al., arXiv:0811.3107, Phys. Rev.
Lett. (2009) in press.

\bibitem{Xiao09}Z.G. Xiao, B.A. Li, L.W.. Chen, G.C. Yong and M. Zhang, Phys. Rev. Lett. {\bf 102}, 062502 (2009).

\bibitem{Rei07} W. Reisdorf et al., Nucl. Phys. A {\bf 781}, 459 (2007).

\bibitem{LiBA00} B.A. Li, Phys. Rev. Lett. {\bf 85}, 4221 (2000).

\bibitem{LiBA02} B.A. Li, Phys. Rev. Lett. {\bf 88}, 192701 (2002); Nucl. Phys.
{\bf A708}, 365 (2002).

\bibitem{Kut94} M. Kutschera, Phys. Lett. {\bf B340}, 1 (1994).

\bibitem{Szm06}A. Szmaglinski et al., Acta Phys. Polon. {\bf B37}, 277 (2006).

\bibitem{Kut00}M. Kutschera et al., Phys. Rev. {\bf C62}, 025802 (2000).

\bibitem{Ban00}S. Banik and D. Bandyopadhyay, J. Phys. G {\bf 26}, 1495 (2000).

\bibitem{BPS71} G. Baym, C. Pethick and P. Sutherland, Astrophys. J. \textbf{170}, 299 (1971).

\bibitem{BBP71} G. Baym, H. A. Bethe and C. J. Pethick, Nucl. Phys. \textbf{A175}, 225 (1971).

\bibitem{Pet95a} C. J. Pethick and D. G. Ravenhall, Ann. Rev. Nucl. Part.
Sci. \textbf{45}, 429 (1995).

\bibitem{Pet95b} C. J. Pethick, D. G. Ravenhall and C. P. Lorenz, Nucl. Phys. \textbf{A584}, 675 (1995).

\bibitem{Lin99} B. Link, R. I. Epstein, J.M. Lattimer, Phys. Rev. Lett. \textbf{83}, 3362 (1999).

\bibitem{Hor04} C.J. Horowitz et al., Phys. Rev. C \textbf{69}, 045804
(2004); C.J. Horowitz et al., Phys. Rev. C \textbf{70}, 065806
(2004)

\bibitem{Bur06} A. Burrows, S. Reddy, and T. A. Thompson, Nucl. Phys. \textbf{A777}, 356 (2006).

\bibitem{Owe05} B. J. Owen, Phys. Rev. Lett. \textbf{95} (2005) 211101.

\bibitem{Rus06} S. B. Ruster, M. Hempel, and J. Schaffner-Bielich, Phys. Rev. C \textbf{73}, 035804 (2006).

\bibitem{Xu09} J. Xu, L.W. Chen, B.A. Li, H.R. Ma, 2008, arXiv:0807.4477v1, Phys. Rev. C (2009) in press; arXiv:0901.2309v1 [astro-ph]

\bibitem{Rav83} D.G. Ravenhall, C.J. Pethick, and J.R. Wilson, Phys. Rev.
Lett. \textbf{50}, 2066 (1983).

\bibitem{Oya93} K. Oyamatsu, Nucl. Phys. \textbf{A561}, 431 (1993).

\bibitem{Ste08} A.W. Steiner, Phys. Rev. C \textbf{77}, 035805 (2008).

\bibitem{Dou00} F. Douchin and P. Haensel, Phys. Lett. \textbf{B485}, 107
(2000).

\bibitem{Dou01} F. Douchin and P. Haensel, A\&A \textbf{380}, 151 (2001).

\bibitem{Hor03} J. Carriere, C.J. Horowitz, and J. Piekarewicz, Astrophys. J. \textbf{593}, 463 (2003).

\bibitem{Oya07} K. Oyamatsu and K. Iida, Phys. Rev. C \textbf{75}, 015801
(2007).

\bibitem{Duc07} C. Ducoin, Ph. Chomaz and F. Gulminelli, Nucl. Phys. \textbf{%
A789}, 403 (2007).

\bibitem{Kub07} S. Kubis, Phys. Rev. C \textbf{76}, 035801 (2007); Phys.
Rev. C \textbf{70}, 065804 (2004).

\bibitem{Wor08} A. Worley, P.G. Krastev and B.A. Li, Astrophys. J. \textbf{%
685}, 390 (2008).

\bibitem{Hor01} C.J. Horowitz and J. Piekarewicz, Phys. Rev. Lett. \textbf{86%
}, 5647 (2001); Phys. Rev. C \textbf{64}, 062802 (R) (2001); Phys.
Rev. C \textbf{66}, 055803 (2002).

\bibitem{Arp72} J. Arponen, Nucl. Phys. \textbf{A191}, 257 (1972).

\bibitem{Ava08b} S.S. Avancini et al., 2008, arXiv:0812.3170v1 [nucl-th]

\bibitem{Maggiore:2007}M.~Maggiore, Nature {\bf 447} (2007)
651.

\bibitem{Flanagan:2005yc}
  E.~E.~Flanagan and S.~A.~Hughes, New J.\ Phys.\  {\bf 7} (2005) 204.

\bibitem{Abbott:2004ig}
  B.~Abbott {\it et al.}  [LIGO Scientific Collaboration],
  Phys.\ Rev.\ Lett.\  {\bf 94} (2005) 181103; Phys.\ Rev.\  D {\bf 76} (2007)
  042001.

\bibitem{Acernese:2007zzb}
  F.~Acernese {\it et al.}, Class.\ Quant.\ Grav.\  {\bf 24} (2007) S491.

\bibitem{Kra08b}P.G. Krastev, B.A. Li and A. Worley, Phys. Lett. {\bf B668}, 1 (2008).

\bibitem{Wor09} A. Worley, P.G. Krastev and B.A., Li, arXiv:0812.0408.

\bibitem{Wen09}D.H. Wen, B.A. Li and P.G. Krastev, preprint (2009).

\bibitem{Kra08a}P.G. Krastev, B.A. Li and A. Worley, ApJ, {\bf 676},
1170 (2008).

\bibitem{HAJS:2007PRL}B. Haskell, N. Andersson, D. I. Jones, and L.
Samuelsson, Phys. Rev. Lett. {\bf 99} (2007) 231101.

\bibitem{Owen:2005PRL}B. J. Owen, Phys. Rev. Lett. {\bf 95} (2005) 211101.

\bibitem{Lattimer:2005}J. M. Lattimer and B. F. Schutz, Astrophys. J. {\bf 629} (2005)
979.
\bibitem{Cha91}S. Chandrasekhar and V. Ferrari, Proc. R. Soc. London A \textbf{432}, 247(1991); {\it ibid}, \textbf{434}, 449(1991).

\bibitem{Nil98}N. Andersson and K.D. Kokkotas,MNRAS, 299, 1059-1068 (1998).

\bibitem{Ben99}O. Benhar, E. Berti and V. Ferrari, MNRAS \textbf{310}, 797(1999).

\bibitem{Tsu05}L.K. Tsui and P.T. Leung, MNRAS \textbf{357}, 1029(2005); Astrophys. J \textbf{631}, 495(2005).
\end{thebibliography}
\end{document}